\documentclass{article}
\usepackage{amssymb, amsthm}

\newtheorem{theorem}{Theorem}

\newtheorem{corollary}{Corollary}

\usepackage{amsmath}
\usepackage{graphicx}

\usepackage{latexsym}

\title{State complexity of union and intersection combined with star and reversal\thanks{All correspondence should be directed to Yuan Gao at
ygao72@csd.uwo.ca. This work is supported by Natural Science and
Engineering Council of Canada Discovery Grant 41630.}}
\author{Yuan Gao, Sheng Yu\\
Department of Computer Science, \\
The University of Western Ontario,\\
London, Ontario, Canada N6A 5B7}
\date{}

\begin{document}
\maketitle

\begin{abstract}
In this paper, we study the state complexities of union and
intersection combined with star and reversal, respectively. We
obtain the state complexities of these combined operations on
regular languages and show that they are less than the mathematical
composition of the state complexities of their individual
participating operations.
\end{abstract}

\section{Introduction}

State complexity is one of the fundamental topics in automata
theory. It is important from both theoretical aspect and
implications in automata applications, because the state complexity
of an operation gives an upper bound of both time and space
complexity of the operation. For example, programmers should know
the largest possible number of states that would be generated before
they perform an operation in an application, since they need to
allocate enough space for the computation and make an estimate of
the time it takes.

The research on state complexity can be recalled to
1950's~\cite{RaSc59}. However, most results on state complexity came
out after
1990~\cite{CCSY99,CaSaYu02,DaDoSa08,Domaratzki02,HoKu02,JiJiSz05,Jriaskova05,JiOk05,PiSh2002,SaWoYu04,Yu01,YuZhSa94}.
Their research focused on individual operations, e.g. union,
intersection, star, catenation, reversal, etc, until A. Salomaa, K.
Salomaa and S. Yu initiated the study of state complexities of
combined operations in 2007~\cite{SaSaYu07}. In the following three
years, many papers were published on this
topic~\cite{CGKY10-cat-sr,CGKY10-cat-ui,DoOk09,EsGaLiYu09,GaSaYu08,GaYu09,JiOk07,LiMaSaYu08}.

People are interested in state complexities of combined operations
not only because it is a relatively new research direction but also
because its importance in practice. For example, several operations
are often applied in a certain order on languages in searching and
language processing. If we simply use the mathematical composition
of the state complexities of individual participating operations, we
may get a very huge value which is far greater than the exact state
complexity of the combined operation, because the resulting
languages of the worst case of one operation may not be among the
worst case input languages of the next
operation~\cite{GaSaYu08,JiOk07,LiMaSaYu08,SaSaYu07}. Although
computer technology is developing fast, time and space should still
be used efficiently. Thus, state complexities of combined operations
are at least as important as those of individual operations.

In~\cite{SaSaYu07}, two combined operations were investigated:
$(L(M)\cup L(N))^*$ and $(L(M)\cap L(N))^*$, where $M$ and $N$ are
$m$-state and $n$-state DFAs, respectively. In~\cite{LiMaSaYu08},
Boolean operations combined with reversal were studied, including:
$(L(M)\cup L(N))^R$ and $(L(M)\cap L(N))^R$. One natural question is
what are the state complexities of these combined operations if we
exchanged the orders of the composed individual operations. For
example, we perform star or reversal first and then perform union or
intersection. Thus, in this paper, we investigate four particular
combined operations: $L(M)^*\cup L(N)$, $L(M)^*\cap L(N)$,
$L(M)^R\cup L(N)$ and $L(M)^R\cap L(N)$.

It has been shown in~\cite{YuZhSa94} that, (1) the state
complexities of the union and intersection of an $m$-state DFA
language and an $n$-state DFA language are both $mn$, (2) the state
complexity of star of a $k$-state DFA language is $\frac{3}{4}2^k$,
and (3), the state complexity of reversal of an $l$-state DFA
language is $2^l$. In this paper, we obtain the state complexities
of $L(M)^*\cup L(N)$, $L(M)^*\cap L(N)$, $L(M)^R\cup L(N)$ and
$L(M)^R\cap L(N)$ and show that they are all less than the
mathematical compositions of individual state complexities for
$m,n\ge 2$.

We prove that the state complexity of $L(M)^*\cup L(N)$ is
$\frac{3}{4}2^m\cdot n-n+1$ for $m$, $n\ge 2$ which is much less
than the known state complexity of $(L(M)\cup L(N))^*$
(\cite{SaSaYu07}). We obtain that the state complexity of
$L(M)^*\cap L(N)$ is also $\frac{3}{4}2^m\cdot n-n+1$ for $m$, $n\ge
2$ whereas the state complexity of $(L(M)\cap L(N))^*$ has been
proved to be $\frac{3}{4}2^{mn}$, the mathematical compositions of
individual state complexities (\cite{SaSaYu07}). For $L(M)^R\cup
L(N)$ and $L(M)^R\cap L(N)$, we prove both of their state
complexities to be $2^m\cdot n-n+1$ for $m$, $n\ge 2$ while the
state complexities of $(L(M)\cup L(N))^R$ and $(L(M)\cap L(N))^R$
are both $2^{m+n}-2^m-2^n+2$ (\cite{LiMaSaYu08}).

In the next section, we introduce the basic notations and
definitions used in this paper. In
Sections~\ref{star-union},~\ref{star-intersection},~\ref{reversal-union}
and~\ref{reversal-intersection}, we investigate the state
complexities of $L(M)^*\cup L(N)$, $L(M)^*\cap L(N)$, $L(M)^R\cup
L(N)$ and $L(M)^R\cap L(N)$, respectively. In
Section~\ref{sec:conclusion}, we conclude the paper .


\section{Preliminaries}
An alphabet $\Sigma$ is a finite set of letters. A word $w \in
\Sigma^*$ is a sequence of letters in $\Sigma$, and the empty word,
denoted by $\varepsilon$, is the word of length 0.

A {\it deterministic finite automaton} (DFA) is usually denoted by a
5-tuple $A = (Q, \Sigma, \delta, s, F)$, where $Q$ is the finite and
nonempty set of states, $\Sigma$ is the finite and nonempty set of
input symbols, $\delta: Q\times\Sigma \rightarrow Q$ is the state
transition function, $s\in Q$ is the initial state, and $F\subseteq
Q$ is the set of final states. A DFA is said to be {\it complete} if
$\delta$ is a total function. Complete DFAs are the basic model for
considering state complexity. Without specific mentioning, all DFAs
are assumed to be complete in this paper. We extend $\delta$ to $Q
\times \Sigma^* \rightarrow Q$ in the usual way. Then this automaton
accepts a word $w \in \Sigma^*$ if $\delta(s,w) \cap F \neq
\emptyset$. Two states in a DFA are said to be {\it equivalent} if
and only if for every word $w \in \Sigma^*$, if $A$ is started in
either state with $w$ as input, it either accepts in both cases or
rejects in both cases. The language accepted by a DFA $A$ is denoted
by $L(A)$. A language is accepted by many DFAs but there is only one
essentially unique {\it minimal} DFA for the language which has the
minimum number of states.

A {\it non-deterministic finite automaton} (NFA) is also denoted by
a 5-tuple $B = (Q, \Sigma, \delta, s, F)$, where $Q$, $\Sigma$, $s$,
and $F$ are defined the same way as in a DFA and $\delta:
Q\times\Sigma\rightarrow 2^Q$ maps a pair consisting of a state and
an input symbol into a set of states rather than a single state. An
NFA may have multiple initial states, in which case an NFA is
denoted $(Q, \Sigma, \delta, S, F)$ where $S$ is the set of initial
states. A language $L$ is accepted by an NFA if and only if $L$ is
accepted by a DFA, and such a language is called a {\it regular
language}. Two finite automata are said to be equivalent if they
accepts the same regular language. An NFA can always be transformed
into an equivalent DFA by performing subset construction. The reader
may refer to~\cite{HoMoUl01,Yu97} for more details about regular
languages and automata theory.

The {\it state complexity} of a regular language $L$ is the number
of states of the minimal, complete DFA accepting $L$. The state
complexity of a class of regular languages is the worst among the
state complexities of all the languages in the class. The state
complexity of an operation on regular languages is the state
complexity of the resulting languages from the operation. For
example, we say that the state complexity of union of an $m$-state
DFA language and an $n$-state DFA language is $mn$. This implies
that the largest number of states of all the minimal, complete DFAs
that accept the union of an $m$-state DFA language and an $n$-state
DFA language,
is $mn$, and such languages exist. Thus, state complexity is a
worst-case complexity.

\section{State complexity of $L_1^*\cup L_2$}\label{star-union}
We first consider the state complexity of $L_1^*\cup L_2$, where
$L_1$ and $L_2$ are regular languages accepted by $m$-state and
$n$-state DFAs, respectively. It has been proved that the state
complexity of $L_1^*$ is $\frac{3}{4}2^m$ and the state complexity
of $L_1\cup L_2$ is $mn$~\cite{Maslov70,YuZhSa94}. The mathematical
composition of them is $\frac{3}{4}2^m\cdot n$. In the following, we
show that this upper bound can be lower.

\begin{theorem}
\label{star union upper bound}

For any $m$-state DFA $M=(Q_M,\Sigma , \delta_M , s_M, F_M)$ and
$n$-state DFA $N=(Q_N,\Sigma , \delta_N , s_N, F_N)$ such that
$|F_M-\{ s_M \}|=k\geq 1$, $m\geq 2$, $n\geq 1$, there exists a DFA
of at most $(2^{m-1}+2^{m-k-1})\cdot n-n+1$ states that accepts
$L(M)^*\cup L(N)$.
\end{theorem}

{\bf Proof.\ \ } Let $M=(Q_M,\Sigma , \delta_M , s_M, F_M)$ be a
complete DFA of $m$ states. Denote $|F_M-\{ s_M \}|$ by $F_0$. Then
$F_0=k\geq 1$ Let $N=(Q_N,\Sigma , \delta_N , s_N, F_N)$ be another
complete DFA of $n$ states. Let DFA $M'=(Q_{M'},\Sigma , \delta_{M'}
, s_{M'}, F_{M'})$ where
\begin{eqnarray*}
& & s_{M'} \notin Q_M\mbox{ is a new start state,}\\
& & Q_{M'} = \{s_{M'}\}\cup \{P\mid P\subseteq (Q_M-F_0)\mbox{ \& } P\neq \emptyset \} \\
& & \qquad \cup \{R\mid R\subseteq Q_M \mbox{ \& } s_M\in R \mbox{ \& }R\cap F_0\neq \emptyset \},\\
& & \delta_{M'}(s_{M'}, a)= \{\delta_M(s_M, a)\mbox{ for any $a\in \Sigma$} \},\\
& & \delta_{M'}(R, a)= \{\delta_M(R, a)\}\mbox{ for $R\subseteq Q_M$ and $a\in \Sigma$ if $\delta_M(R, a)\cap F_0=\emptyset$},\\
& & \delta_{M'}(R, a)= \{ \delta_M(R, a)\}\cup \{s_M\}\mbox{ otherwise}, \\
& & F_{M'}= \{s_{M'}\}\cup\{R\mid R\subseteq Q_M \mbox{ \& } R\cap
F_M\neq \emptyset \}.
\end{eqnarray*}

It is clear that $M'$ accepts $L(M)^*$. In the second term of the
union for $Q_{M'}$ there are $2^{m-k}-1$ states. And in the third
term, there are $(2^k-1)2^{m-k-1}$ states. So $M'$ has
$2^{m-1}+2^{m-k-1}$ states in total. Now we construct another DFA
$A=(Q,\Sigma , \delta , s, F)$ where
\begin{eqnarray*}
& & s=\langle s_{M'},s_N \rangle,\\
& & Q = \{\langle i,j \rangle \mid i\in Q_{M'}-\{s_{M'}\},j\in Q_N\}\cup \{s \}, \\
& & \delta(\langle i,j \rangle, a)= \langle \delta_{M'}(i, a),\delta_N(j, a) \rangle \mbox{, $\langle i,j \rangle \in Q$, $a\in \Sigma$},\\
& & F= \{\langle i,j \rangle \mid i\in F_{M'}\mbox{ or }j\in F_N \}.
\end{eqnarray*}
We can see that $$L(A)=L(M')\cup L(N)=L(M)^*\cup L(N).$$ Note
$\langle s_{M'},j \rangle \notin Q$, for $j\in Q_N-\{s_N\}$, because
there is no transition going into $s_{M'}$ in DFA $M'$. So there are
at least $n-1$ states in $Q$ are not reachable. Thus, the number of states of minimal DFA accepting $L(M)^*\cup L(N)$ is no more than\\
$$|Q|=(2^{m-1}+2^{m-k-1})\cdot n-n+1. \ \ \Box$$

If $s_M$ is the only final state of $M$($k=0$), then $L(M)^*=L(M)$.

\begin{corollary}
\label{star union upper bound corollary}

For any $m$-state DFA $M=(Q_M,\Sigma , \delta_M , s_M, F_M)$ and
$n$-state DFA $N=(Q_N,\Sigma , \delta_N , s_N, F_N)$, $m>1$, $n>0$,
there exists a DFA $A$ of at most $\frac{3}{4}2^m\cdot n-n+1$ states
such that $L(A)=L(M)^*\cup L(N)$.
\end{corollary}
{\bf Proof.\ \ } Let $k$ be defined as in the above proof. There are
two cases in the following.
\begin{itemize}
\item[{\rm (I)}]$k=0$. In this case, $L(M)^*=L(M)$. Then $A$ simply needs at most $m\cdot
n$ states, which is less than $\frac{3}{4}2^m\cdot n-n+1$ when
$m>1$.
\item[{\rm (II)}]$k\geq 1$. The claim is clearly true by Theorem~\ref{star union upper bound}.
$\ \ \Box$
\end{itemize}
Next, we show that the upper bound $\frac{3}{4}2^m\cdot n-n+1$ is
reachable.
\begin{theorem}
\label{star union lower bound}

Given two integers $m\geq 2$, $n\geq 2$, there exists a DFA $M$ of
$m$ states and a DFA $N$ of $n$ states such that any DFA accepting
$L(M)^*\cup L(N)$ needs at least $\frac{3}{4}2^m\cdot n-n+1$ states.
\end{theorem}

\begin{figure}[ht]
  \centering
  \includegraphics[scale=0.60]{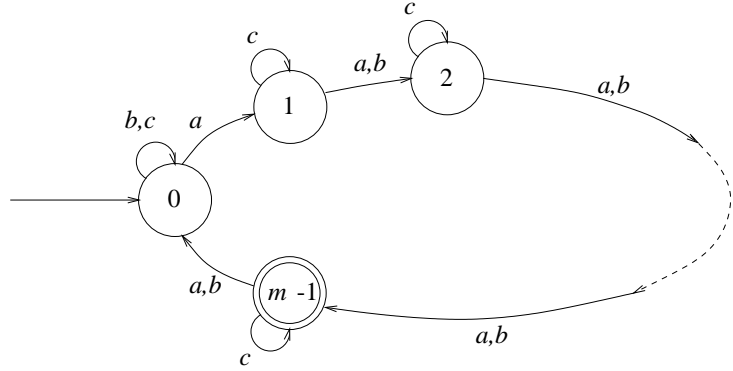}
  \caption{The transition diagram of the witness DFA $M$ of Theorems~\ref{star union lower bound} and~\ref{star intersection lower bound}}
\label{star-union-first}
\end{figure}

\begin{figure}[ht]
  \centering
  \includegraphics[scale=0.60]{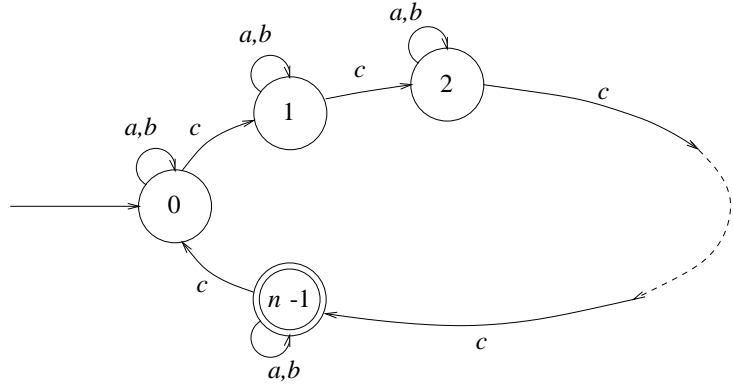}
  \caption{The transition diagram of the witness DFA $N$ of Theorems~\ref{star union lower bound} and~\ref{star intersection lower bound}}
\label{star-union-second}
\end{figure}

{\bf Proof.\ \ } Let $M=(Q_M,\Sigma , \delta_M , 0, \{ m-1 \})$ be a
DFA, where $Q_M = \{0,1,\ldots ,m-1\}$, $\Sigma = \{a,b,c\}$ and the
transitions of $M$ are
\begin{eqnarray*}
& & \delta_M(i, a) = i+1 \mbox{ mod $m$, } i=0,1, \ldots , m-1,\\
& & \delta_M(0, b) = 0 \mbox{, }\delta_M(i, b) = i+1 \mbox{ mod $m$, } i=1, \ldots , m-1,\\
& & \delta_M(i, c) = i \mbox{, } i=0,1, \ldots , m-1.
\end{eqnarray*}
The transition diagram of $M$ is shown in
Figure~\ref{star-union-first}.
\\
Let $N=(Q_N,\Sigma , \delta_N , 0, \{n-1\})$ be another DFA, where
$Q_N = \{0,1,\ldots ,n-1\}$ and
\begin{eqnarray*}
& & \delta_N(i, a) = i \mbox{, } i=0,1, \ldots , n-1,\\
& & \delta_N(i, b) = i \mbox{, } i=0,1, \ldots , n-1,\\
& & \delta_N(i, c) = i+1 \mbox{ mod $n$, } i=0,1, \ldots , n-1.
\end{eqnarray*}
The transition diagram of $N$ is shown in
Figure~\ref{star-union-second}.


It has been proved in~\cite{YuZhSa94} that the minimal DFA accepting
the star of an $m$-state DFA language has $\frac{3}{4}2^m$ states in
the worst case. $M$ is a modification of worst case example given in
~\cite{YuZhSa94} by adding a $c$-loop to every state. So we design a
$\frac{3}{4}2^m$-state, minimal DFA $M'=(Q_{M'},\Sigma , \delta_{M'}
, s_{M'}, F_{M'})$ that accepts $L(M)^*$, where
\begin{eqnarray*}
& & s_{M'} \notin Q_M\mbox{ is a new start state,}\\
& & Q_{M'} = \{s_{M'}\}\cup \{P\mid P\subseteq \{0,1,\ldots ,m-2\}\mbox{ \& } P\neq \emptyset \} \\
& & \qquad \cup \{R\mid R\subseteq \{0,1,\ldots ,m-1\} \mbox{ \& } 0\in R \mbox{ \& }m-1\in R \},\\
& & \delta_{M'}(s_{M'}, a)= \{\delta_M(0, a)\mbox{ for any $a\in \Sigma$} \},\\
& & \delta_{M'}(R, a)= \{\delta_M(R, a)\}\mbox{ for $R\subseteq Q_M$ and $a\in \Sigma$ if $m-1\notin \delta_M(R, a)$},\\
& & \delta_{M'}(R, a)= \{ \delta_M(R, a)\}\cup \{0\}\mbox{ otherwise}, \\
& & F_{M'}= \{s_{M'}\}\cup\{R\mid R\subseteq \{0,1,\ldots ,m-1\}
\mbox{ \& } m-1\in R\}.
\end{eqnarray*}

Then we construct a DFA $A=(Q,\Sigma , \delta , s, F)$ accepting
$L(M)^*\cup L(N)$ exactly as described in the proof of
Theorem~\ref{star union upper bound}, where
\begin{eqnarray*}
& & s=\langle s_{M'},0 \rangle,\\
& & Q = \{\langle i,j \rangle \mid i\in Q_{M'}-\{s_{M'}\},j\in Q_N\}\cup \{s \}, \\
& & \delta(\langle i,j \rangle, a)= \langle \delta_{M'}(i, a),\delta_N(j, a) \rangle \mbox{, $\langle i,j \rangle \in Q$, $a\in \Sigma$},\\
& & F= \{\langle i,j \rangle \mid i\in F_{M'}\mbox{ or }j=n-1 \}.
\end{eqnarray*}

Now we need to show that $A$ is a minimal DFA.
\begin{itemize}
\item[{\rm (I)}]All the states in $Q$ are reachable.\\
For an arbitrary state $\langle i,j\rangle$ in $Q$, there always
exists a string $w_1w_2$ such that $\delta(\langle s_M',0\rangle,
w_1w_2) = \langle i,j\rangle$, where
\begin{eqnarray*}
& & \delta_{M'}(s_{M'}, w_1)=i\mbox{, }w_1\in \{a,b\}^*,\\
& & \delta_N (0, w_2)=j\mbox{, }w_2\in \{c\}^*.
\end{eqnarray*}
\item[{\rm (II)}]Any two different states $\langle i_1,j_1\rangle$ and $\langle i_2,j_2\rangle$ in $Q$ are
distinguishable.\\
\begin{itemize}
\item[{\rm 1.}]$i_1\neq i_2$, $j_2\neq n-1$. We can find a string $w_1$ such that
\begin{eqnarray*}
& & \delta(\langle i_1,j_1\rangle, w_1)\in F,\\
& & \delta(\langle i_2,j_2\rangle, w_1) \notin F,
\end{eqnarray*}
where $w_1\in \{a,b\}^*$, $\delta_{M'}(i_1, w_1)\in  F_{M'}$ and
$\delta_M'(i_2, w_1) \notin F_M'$.

\item[{\rm 2.}]$i_1\neq i_2$, $j_2= n-1$. There exists a string $w_1$ such that
\begin{eqnarray*}
& & \delta(\langle i_1,j_1\rangle, w_1c)\in F,\\
& & \delta(\langle i_2,j_2\rangle, w_1c) \notin F,
\end{eqnarray*}
where $w_1\in \{a,b\}^*$, $\delta_{M'}(i_1, w_1)\in  F_{M'}$ and
$\delta_{M'}(i_2, w_1) \notin F_{M'}$.
\item[{\rm 3.}]$i_1= i_2\notin F_{M'}$, $j_1\neq j_2$. For this case, a string $c^{n-1-j_1}$ can distinguish the two states, since $\delta(\langle i_1,j_1\rangle, c^{n-1-j_1})\in
F$ and $\delta(\langle i_2,j_2\rangle, c^{n-1-j_1}) \notin F$.

\item[{\rm 4.}]$i_1= i_2\in F_{M'}$, $j_1\neq j_2$. A string $b^mc^{n-1-j_1}$ can distinguish them, because $\delta(\langle i_1,j_1\rangle, b^mc^{n-1-j_1})\in
F$ and $\delta(\langle i_2,j_2\rangle, b^mc^{n-1-j_1}) \notin F$.

\end{itemize}
\end{itemize}
Since all the states in $A$ are reachable and distinguishable, DFA
$A$ is minimal. Thus, any DFA accepting $L(M)^*\cup L(N)$ needs at
least $\frac{3}{4}2^m\cdot n-n+1$ states. $\ \ \Box $

This result gives a lower bound for the state complexity of
$L(M)^*\cup L(N)$. It coincides with the upper bound in
Corollary~\ref{star union upper bound corollary}. So we have the
following Theorem~\ref{Tight bound of star union}.
\begin{theorem}
\label{Tight bound of star union}

For any integer $m\geq 2$, $n\geq 2$, $\frac{3}{4}2^m\cdot n-n+1$
states are both sufficient and necessary in the worst case for a DFA
to accept $L(M)^*\cup L(N)$, where $M$ is an $m$-state DFA and $N$
is an $n$-state DFA.
\end{theorem}

\section{State complexity of $L(M)^*\cap L(N)$}\label{star-intersection}

Since the state complexity of intersection on regular languages is
the same as that of union~\cite{YuZhSa94}, the mathematical
composition of the state complexities of star and intersection is
also $\frac{3}{4}2^m$. In this section, we show that the state
complexity of $L(M)^*\cap L(N)$ is $\frac{3}{4}2^m\cdot n-n+1$ which
is the same as the state complexity of $L(M)^*\cup L(N)$.

\begin{theorem}
\label{star intersection upper bound}

For any $m$-state DFA $M=(Q_M,\Sigma , \delta_M , s_M, F_M)$ and
$n$-state DFA $N=(Q_N,\Sigma , \delta_N , s_N, F_N)$ such that
$|F_M-\{ s_M \}|=k\geq 1$, $m>1$, $n>0$, there exists a DFA of at
most $(2^{m-1}+2^{m-k-1})\cdot n-n+1$ states that accepts
$L(M)^*\cap L(N)$.
\end{theorem}

{\bf Proof.\ \ } We construct a DFA $A$ accepting $L(M)^*\cap L(N)$
the same as in the proof of Theorem~\ref{star union upper bound}
except that its set of final states is
\[
F= \{\langle i,j \rangle \mid i\in F_{M'}\mbox{, }j\in F_N \}.
\]
Thus, after reducing the $n-1$ unreachable states $\langle s_{M'},j
\rangle \notin Q$, for $j\in Q_N-\{s_N\}$, the number of states of
$A$ is sill no more than $(2^{m-1}+2^{m-k-1})\cdot n-n+1. \ \ \Box$

Similarly to the proof of Corollary~\ref{star union upper bound
corollary}, we consider both the case that $M$ has no other final
state except $s_M$ ($L(M)^*=L(M)$) and the case that $M$ has some
other final states (Theorem~\ref{star intersection upper bound}).
Then we obtain the following corollary. Detailed proof may be
omitted.

\begin{corollary}
\label{star intersection upper bound corollary}

For any $m$-state DFA $M=(Q_M,\Sigma , \delta_M , s_M, F_M)$ and
$n$-state DFA $N=(Q_N,\Sigma , \delta_N , s_N, F_N)$, $m>1$, $n>0$,
there exists a DFA $A$ of at most $\frac{3}{4}2^m\cdot n-n+1$ states
such that $L(A)=L(M)^*\cap L(N)$.
\end{corollary}
Next, we show that this general upper bound of state complexity of
$L(M)^*\cap L(N)$ can be reached by some witness DFAs.
\begin{theorem} \label{star intersection lower bound}

Given two integers $m\geq 2$, $n\geq 2$, there exists a DFA $M$ of
$m$ states and a DFA $N$ of $n$ states such that any DFA accepting
$L(M)^*\cap L(N)$ needs at least $\frac{3}{4}2^m\cdot n-n+1$ states.
\end{theorem}

{\bf Proof.\ \ } We use the same DFAs $M$ and $N$ as in the proof of
Theorem~\ref{star union lower bound}. Their transition diagrams are
shown in Figure~\ref{star-union-first} and
Figure~\ref{star-union-second}, respectively. Construct DFA
$M'=(Q_{M'},\Sigma , \delta_{M'} , s_{M'}, F_{M'})$ that accepts
$L(M)^*$ in the same way.

Then we construct a DFA $A=(Q,\Sigma , \delta , s, F)$ accepting
$L(M)^*\cap L(N)$ exactly as described in the proof of
Theorem~\ref{star union lower bound} except that
\[
F= \{\langle i,n-1 \rangle \mid i\in F_{M'} \}.
\]

Now we prove that $A$ is minimal.
\begin{itemize}
\item[{\rm (I)}]Every state of $A$ is reachable.\\
Let $\langle i,j\rangle$ be an arbitrary state of $A$. Then there
always exists a string $w_1w_2$ such that $\delta(\langle
s_{M'},0\rangle, w_1w_2) = \langle i,j\rangle$, where
\begin{eqnarray*}
& & \delta_{M'}(s_{M'}, w_1)=i\mbox{, }w_1\in \{a,b\}^*,\\
& & \delta_N (0, w_2)=j\mbox{, }w_2\in \{c\}^*.
\end{eqnarray*}
\item[{\rm (II)}]Any two different states $\langle i_1,j_1\rangle$ and $\langle i_2,j_2\rangle$ of $A$ are
distinguishable.\\
\begin{itemize}
\item[{\rm 1.}]$i_1\neq i_2$.

We can find a string $w_1$ such that
\begin{eqnarray*}
& & \delta(\langle i_1,j_1\rangle, w_1c^{n-1-j_1})\in F,\\
& & \delta(\langle i_2,j_2\rangle, w_1c^{n-1-j_1}) \notin F,
\end{eqnarray*}
where $w_1\in \{a,b\}^*$, $\delta_{M'}(i_1, w_1)\in  F_{M'}$ and
$\delta_{M'}(i_2, w_1) \notin F_{M'}$.

\item[{\rm 2.}]$i_1= i_2\notin F_{M'}$, $j_1\neq j_2$.

There exists a string $w_2$ such that
\begin{eqnarray*}
& & \delta(\langle i_1,j_1\rangle, w_2c^{n-1-j_1})\in F,\\
& & \delta(\langle i_2,j_2\rangle, w_2c^{n-1-j_1}) \notin F,
\end{eqnarray*}
where $w_1\in \{a,b\}^*$ and $\delta_{M'}(i_1, w_2)\in  F_{M'}$.
\item[{\rm 3.}]$i_1= i_2\in F_{M'}$, $j_1\neq j_2$.
\begin{eqnarray*}
& & \delta(\langle i_1,j_1\rangle, c^{n-1-j_1})\in F,\\
& & \delta(\langle i_2,j_2\rangle, c^{n-1-j_1}) \notin F.
\end{eqnarray*}
\end{itemize}
\end{itemize}
Due to (I) and (II), $A$ is a minimal DFA with $\frac{3}{4}2^m\cdot
n-n+1$ states which accepts $L(M)^*\cap L(N)$. $\ \ \Box $

This lower bound coincides with the upper bound in
Corollary~\ref{star intersection upper bound corollary}. Thus, the
bounds are tight.
\begin{theorem}
\label{Tight bound of star intersection}

For any integer $m\geq 2$, $n\geq 2$, $\frac{3}{4}2^m\cdot n-n+1$
states are both sufficient and necessary in the worst case for a DFA
to accept $L(M)^*\cap L(N)$, where $M$ is an $m$-state DFA and $N$
is an $n$-state DFA.
\end{theorem}

\section{State complexity of $L_1^R\cup L_2$}\label{reversal-union}
In this section, we study the state complexity of $L_1^R\cup L_2$,
where $L_1$ and $L_2$ are regular languages. It has been proved that
the state complexity of $L_1^R$ is $2^m$ and the state complexity of
$L_1\cup L_2$ is $mn$~\cite{Maslov70,YuZhSa94}. Thus, the
mathematical composition of them is $2^m\cdot n$. In this section we
will prove that this upper bound of state complexity of $L_1^R\cup
L_2$ can not be reached in any case. We will first try to lower the
upper bound in the following.

\begin{theorem}
\label{reversal uion upper bound}

Let $L_1$ and $L_2$ be two regular language accepted by an $m$-state
and $n$-state DFAs, respectively. Then there exists a DFA of at most
$2^m\cdot n-n+1$ states that accepts $L_1^R\cup L_2$.
\end{theorem}

{\bf Proof.\ \ } Let $M=(Q_M,\Sigma , \delta_M , s_M, F_M)$ be a
complete DFA of $m$ states and $L_1=L(M)$. Let $N=(Q_N,\Sigma ,
\delta_N , s_N, F_N)$ be another complete DFA of $n$ states and
$L_2=L(N)$. Let $M'=(Q_M,\Sigma , \delta_{M'} , F_M, \{s_M\})$ be an
NFA with multiple initial states. $\delta_{M'}(p,a)=q$ if
$\delta_M(q,a)=p$ where $a\in \Sigma$ and $p,q\in Q_M$. Clearly,
$L(M')=L(M)^R=L_1^R$. After performing subset construction, we can
get a $2^m$-state DFA $A=(Q_A,\Sigma , \delta_A , s_A, F_A)$ that is
equivalent to $M'$. Since $A$ has $2^m$ states, one of its final
state must be $Q_M$. Now we construct a DFA $B=(Q_B,\Sigma ,
\delta_B , s_B, F_B)$, where
\begin{eqnarray*}
& & Q_B = \{\langle i,j \rangle \mid i\in Q_A\mbox{, } j\in Q_N\},\\
& & s_B = \langle s_A,s_N \rangle,\\
& & F_B = \{\langle i,j \rangle\in Q_B\mid i\in F_A\mbox{ or } j\in F_N\},\\
& & \delta_B(\langle i,j \rangle, a) = \langle i',j' \rangle \mbox{,
if } \delta_A(i,a)=i'\mbox{ and }\delta_N(j,a)=j'\mbox{, }a\in
\Sigma.
\end{eqnarray*}
It is easy to see that $\delta_B(\langle Q_M,j \rangle, a) \in F_B$
for any $j\in Q_N$ and $a\in \Sigma$. This means all the states
(two-tuples) starting with $Q_1$ are equivalent. There are $n$ such
states in total. Thus, the minimal DFA accepting $L_1^R\cup L_2$ has
no more than $2^m\cdot n-n+1$ states.$\ \ \Box $

This result gives an upper bound of state complexity of $L_1^R\cup
L_2$. Now let's see if this bound is reachable.

\begin{theorem}
\label{reversal uion lower bound}

Given two integers $m\geq 2$, $n\geq 2$, there exists a DFA $M$ of
$m$ states and a DFA $N$ of $n$ states such that any DFA accepting
$L(M)^R\cup L(N)$ needs at least $2^m\cdot n-n+1$ states.
\end{theorem}

\begin{figure}[ht]
  \centering
  \includegraphics[scale=0.60]{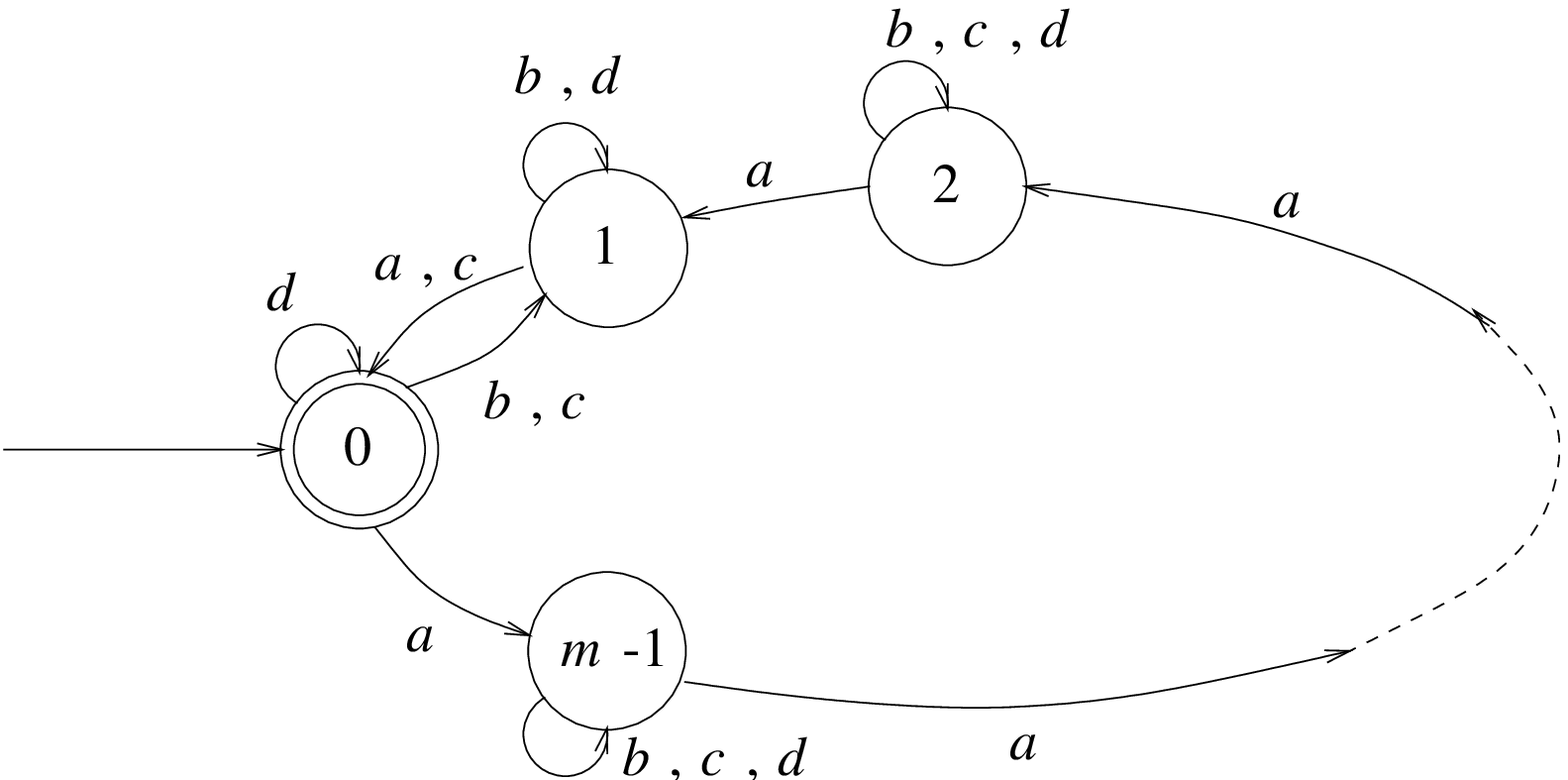}
  \caption{The transition diagram of the witness DFA $M$ of Theorems~\ref{reversal uion lower bound} and~\ref{reversal intersection lower bound}}
\label{reversal-union-first}
\end{figure}

\begin{figure}[ht]
  \centering
  \includegraphics[scale=0.60]{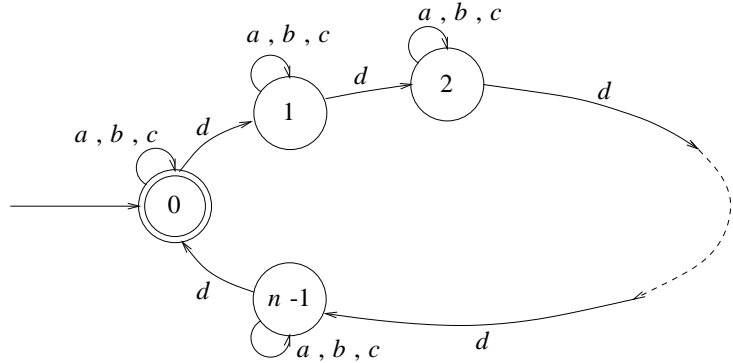}
  \caption{The transition diagram of the witness DFA $N$ of Theorems~\ref{reversal uion lower bound} and~\ref{reversal intersection lower bound}}
\label{reversal-union-second}
\end{figure}

{\bf Proof.\ \ } Let $M=(Q_M,\Sigma , \delta_M , 0, \{0\})$ be a
DFA, where $Q_M = \{0,1,\ldots ,m-1\}$, $\Sigma = \{a,b,c,d\}$ and
the transitions are
\begin{eqnarray*}
& & \delta_M(0, a) = m-1 \mbox{, }\delta_M(i, a) = i-1 \mbox{, } i=1, \ldots , m-1,\\
& & \delta_M(0, b) = 1 \mbox{, } \delta_M(i, b) = i \mbox{, } i=1, \ldots , m-1,\\
& & \delta_M(0, c) = 1 \mbox{, } \delta_M(1, c) = 0 \mbox{, }\delta_M(j, c) = i \mbox{, } j=2, \ldots , m-1,\\
& & \delta_M(k, d) = k \mbox{, } k=0, \ldots , m-1.
\end{eqnarray*}
The transition diagram of $M$ is shown in
Figure~\ref{reversal-union-first}.
Let $N=(Q_N,\Sigma , \delta_N , 0, \{0\})$ be another DFA, where
$Q_N = \{0,1,\ldots ,n-1\}$, $\Sigma = \{a,b,c,d\}$ and the
transitions are
\begin{eqnarray*}
& & \delta_N(i, a) = i \mbox{, } i=0, \ldots , n-1,\\
& & \delta_N(i, b) = i \mbox{, } i=0, \ldots , n-1,\\
& & \delta_N(i, c) = i \mbox{, } i=0, \ldots , n-1,\\
& & \delta_N(i, d) = i+1 \mbox{ mod }n \mbox{, } i=0, \ldots , n-1.
\end{eqnarray*}
The transition diagram of $N$ is shown in
Figure~\ref{reversal-union-second}.

Note that $M$ is a modification of worst case example given
in~\cite{YuZhSa94} for reversal, by adding a $d$-loop to every
state. Intuitively, the minimal DFA accepting $L(M)^R$ should also
have $2^m$ states. Before using this result, we will prove it first.
Let $A=(Q_A,\Sigma , \delta_A , \{0\}, F_A)$ be a DFA, where
\begin{eqnarray*}
& & Q_A = \{q\mid q\subseteq Q_M\},\\
& & \Sigma = \{a,b,c,d\},\\
& & \delta_A(p, e) = \{j\mid \delta_M(i, e)=j\mbox{, }i\in p\} \mbox{, } p\in Q_A\mbox{, } e\in \Sigma,\\
& & F_A = \{q\mid \{0\}\in q \mbox{, }q\in Q_A\}.
\end{eqnarray*}
Clearly, $A$ has $2^m$ states and it accepts $L(M)^R$. Now let's
prove it is minimal.
\begin{itemize}
\item[{\rm (i)}]Every state $i \in Q_A$ is
reachable.\\
\begin{itemize}
\item[{\rm 1.}]$i=\emptyset$.\\
$|i|=0$ if and only if $i=\emptyset$. $\delta_A(\{ 0 \}, b) =
i=\emptyset .$
\item[{\rm 2.}]$|i|=1$.\\
Assume that $i=\{ p \}$, $0\leq p\leq m-1$. $\delta_A(\{ 0 \}, a^p)
=i.$
\item[{\rm 3.}]$2\leq |i|\leq m$.\\
Assume that $i=\{ i_1, i_2, \ldots ,i_k \}$, $0\leq i_1<i_2< \ldots
<i_k \leq m-1$, $2\leq k\leq m$. $\delta_A(\{ 0 \}, w) = i$, where
$$w = ab(ac)^{i_k-i_{k-1}-1}ab(ac)^{i_{k-1}-i_{k-2}-1}\cdots ab(ac)^{i_2-i_1-1}a^{i_1}.$$
\end{itemize}

\item[{\rm (ii)}]Any two different states $i$ and $j$ in $Q_A$ are
distinguishable.\\
Without loss of generality, we may assume that $|i|\geq |j|$. Let
$x\in i-j$. Then a string $a^{m-x}$ can distinguish these two states
because
\begin{eqnarray*}
\delta_A(i, a^{m-x})& \in & F_A,\\
\delta_A(j, a^{m-x}) & \notin & F_A.
\end{eqnarray*}

\end{itemize}
Thus, $A$ is a minimal DFA with $2^m$ states which accepts $L(M)^R$.
Now let $B=(Q_B,\Sigma , \delta_B , \{\langle \{0\},0\rangle\},
F_B)$ be a DFA, where
\begin{eqnarray*}
& & Q_B = \{\langle p,q\rangle \mid p\in Q_A-\{ Q_M\}\mbox{, }q\in Q_N\}\cup \{\langle Q_M,0\rangle \},\\
& & \Sigma = \{a,b,c,d\},\\
& & F_B = \{\langle p,q\rangle\mid p\in F_A \mbox{ or }q\in F_N
\mbox{, }\langle p,q\rangle\in Q_B\},
\end{eqnarray*}
and for $\langle p,q\rangle\in Q_B$, $,e\in \Sigma$
\begin{eqnarray*}
\delta_B(\langle p,q\rangle, e) & = & \left\{
\begin{array}{l l}
  \langle p',q'\rangle & \mbox{if } \delta_A(p, e)=p'\mbox{, } \delta_N(q, e)=q'\mbox{, $p'\neq Q_M$,}\\
  \langle Q_M,0\rangle & \mbox{if } \delta_A(p, e)=Q_M.
\end{array} \right.
\end{eqnarray*}

As we mentioned in last proof, all the states (two-tuples) starting
with $Q_M$ are equivalent. Thus, we replace them with one state:
$\langle Q_M,0\rangle$. It is easy to see that $B$ accepts the
language $L(M)^R\cup L(N)$. It has $2^m\cdot n-n+1$ states. Now lets
see if $B$ is a minimal DFA.
\begin{itemize}
\item[{\rm (I)}]All the states in $Q_B$ are reachable.

For an arbitrary state $\langle p,q\rangle$ in $Q_B$, there always
exists a string $d^qw$ such that $\delta_B(\langle \{0\},0\rangle,
d^qw) = \langle p,q\rangle$, where $w\in \{a,b,c\}^*$ and
$\delta_A(\{0\},w)=p$.
\item[{\rm (II)}]Any two different states $\langle p_1,q_1\rangle$ and $\langle p_2,q_2\rangle$ in $Q_B$ are distinguishable.
\begin{itemize}
\item[{\rm 1.}]$q_1=q_2$.

We can easily find a string $d^iw$ such that
\begin{eqnarray*}
\delta_B(\langle p_1,q_1\rangle, d^iw)& \in & F_B,\\
\delta_B(\langle p_2,q_2\rangle, d^iw) & \notin & F_B,
\end{eqnarray*}
where $i+q_1\mbox{ mod }n\neq 0$, $w\in \{a,b,c\}^*$, $\delta_A(p_1,
w)\in F_A$ and $\delta_A(p_2, w) \notin F_A$.
\item[{\rm 2.}]$p_1=p_2$, $q_1\neq q_2$.

A string $d^{n-q_1}w$ can distinguish these two states where $w\in
\{a,b,c\}^*$ and $\delta_A(p_1, w)\notin F_A$, because
\begin{eqnarray*}
\delta_B(\langle p_1,q_1\rangle, d^{n-q_1}w)& \in & F_B,\\
\delta_B(\langle p_2,q_2\rangle, d^{n-q_1}w) & \notin & F_B.
\end{eqnarray*}
\item[{\rm 3.}]$p_1\neq p_2$, $q_1\neq q_2$.

We first find a string $w\in \{a,b,c\}^*$ such that $\delta_A(p_1,
w)\in F_A$ and $\delta_A(p_2, w) \notin F_A$. Then it is clear that
\begin{eqnarray*}
\delta_B(\langle p_1,q_1\rangle, d^{n-q_1}w)& \in & F_B,\\
\delta_B(\langle p_2,q_2\rangle, d^{n-q_1}w) & \notin & F_B.
\end{eqnarray*}
\end{itemize}
\end{itemize}
Since all the states in $B$ are reachable and distinguishable, DFA
$B$ is minimal. Thus, any DFA accepting $L(M)^R\cup L(N)$ needs at
least $2^m\cdot n-n+1$ states. $\ \ \Box $

This result gives a lower bound for the state complexity of
$L(M)^R\cup L(N)$. It coincides with the upper bound. So we have the
following Theorem~\ref{Tight bound of reversal union}.
\begin{theorem}
\label{Tight bound of reversal union}

For any integer $m\geq 2$, $n\geq 2$, $2^m\cdot n-n+1$ states are
both sufficient and necessary in the worst case for a DFA to accept
$L(M)^R\cup L(N)$, where $M$ is an $m$-state DFA and $N$ is an
$n$-state DFA.
\end{theorem}

\section{State complexity of $L_1^R\cap L_2$}\label{reversal-intersection}
The mathematical composition of the state complexities of reversal
and intersection is also $2^m\cdot n$, since the state complexities
of intersection and union are the same~\cite{YuZhSa94}. In this
section, we will show that the state complexity of $L_1^R\cap L_2$
is also $2^m\cdot n-n+1$, which is the same as that of $L_1^R\cup
L_2$. We will start with an upper bound less than the mathematical
composition.

\begin{theorem}
\label{reversal intersection upper bound}

Let $L_1$ and $L_2$ be two regular language accepted by an $m$-state
and $n$-state DFAs, respectively. Then there exists a DFA of at most
$2^m\cdot n-n+1$ states that accepts $L_1^R\cap L_2$.
\end{theorem}

{\bf Proof.\ \ } Let $M=(Q_M,\Sigma , \delta_M , s_M, F_M)$ be a
complete DFA of $m$ states and $L_1=L(M)$. Let $N=(Q_N,\Sigma ,
\delta_N , s_N, F_N)$ be another complete DFA of $n$ states and
$L_2=L(N)$. Let $M'=(Q_M,\Sigma , \delta_{M'} , F_M, \{s_M\})$ be an
NFA with multiple initial states. $\delta_{M'}(p,a)=q$ if
$\delta_M(q,a)=p$ where $a\in \Sigma$ and $p,q\in Q_M$. Clearly,
$L(M')=L(M)^R=L_1^R$. After performing subset construction, we can
get a $2^m$-state DFA $A=(Q_A,\Sigma , \delta_A , s_A, F_A)$ that is
equivalent to $M'$. Since $A$ has $2^m$ states, one of its nonfinal
state must be a sink state, denoted by $t_A$. Now we construct a DFA
$B=(Q_B,\Sigma , \delta_B , s_B, F_B)$, where
\begin{eqnarray*}
Q_B & = & \{\langle i,j \rangle \mid i\in Q_A\mbox{, } j\in Q_N\},\\
s_B & = & \langle s_A,s_N \rangle,\\
F_B & = & \{\langle i,j \rangle\in Q_B \mid i\in F_A\mbox{, } j\in F_N\},\\
\delta_B(\langle i,j \rangle, a) & = & \langle i',j' \rangle \mbox{,
if } \delta_A(i,a)=i'\mbox{ and }\delta_N(j,a)=j'\mbox{, }a\in
\Sigma.
\end{eqnarray*}
We can see that $\delta_B(\langle t_A, j \rangle, a) \notin F_B$ for
any $j\in Q_N$ and $a\in \Sigma$, since $t_A$ is the sink state of
DFA $A$ which accepts $L(M)^R$. This means all the states
(two-tuples) starting with $t_A$ are equivalent. There are $n$ such
states in total. Thus, after reducing them to one state, we can see
the number of states of $A$ is sill no more than $2^m\cdot n-n+1$.
$\ \ \Box$

Theorem~\ref{reversal intersection upper bound} gives an upper bound
of state complexity of $L_1^R\cap L_2$. Now let's see if this bound
is reachable.

\begin{theorem}
\label{reversal intersection lower bound} Given two integers $m\geq
2$, $n\geq 2$, there exists a DFA $M$ of $m$ states and a DFA $N$ of
$n$ states such that any DFA accepting $L(M)^R\cap L(N)$ needs at
least $2^m\cdot n-n+1$ states.
\end{theorem}

{\bf Proof.\ \ } We use the same DFAs $M$ and $N$ as in the proof of
Theorem~\ref{reversal uion lower bound}. Their transition diagrams
are shown in Figure~\ref{reversal-union-first} and
Figure~\ref{reversal-union-second}, respectively. It has been shown
in the proof of Theorem~\ref{reversal uion lower bound} that the
minimal DFA accepting $L(M)^R$ has $2^m$ states. So we design a
minimal DFA $A=(Q_A,\Sigma , \delta_A , \{0\}, F_A)$ that accepts
$L(M)^R$ in the same way, where
\begin{eqnarray*}
Q_A & = & \{q \mid q\subseteq Q_M\},\\
\Sigma & = & \{a,b,c,d\},\\
\delta_A(p, e) & = & \{j\mid \delta_M(i, e)=j\mbox{, }i\in p\} \mbox{, } p\in Q_A\mbox{, } e\in \Sigma,\\
F_A & = & \{q \mid \{0\}\in q \mbox{, }q\in Q_A\}.
\end{eqnarray*}. Note that $A$ must have a sink state,
denoted by $t_A$.

Next we construct a DFA $B=(Q_B,\Sigma , \delta_B , \langle
\{0\},0\rangle, F_B)$ accepting $L(M)^R\cap L(N)$ , where
\begin{eqnarray*}
Q_B & = & \{\langle p,q\rangle \mid p\in Q_A-\{ t_A\}\mbox{, }q\in Q_N\}\cup \{\langle t_A,0\rangle \},\\
\Sigma & = & \{a,b,c,d\},\\
F_B & = & \{\langle p,q\rangle \mid p\in F_A \mbox{, }q\in F_N
\mbox{, }\langle p,q\rangle\in Q_B\},
\end{eqnarray*}
and for $\langle p,q\rangle\in Q_B$, $,e\in \Sigma$
\begin{eqnarray*}
\delta_B(\langle p,q\rangle, e) & = & \left\{
\begin{array}{l l}
  \langle p',q'\rangle & \mbox{if } \delta_A(p, e)=p'\mbox{, } \delta_N(q, e)=q'\mbox{, $p'\neq t_A$,}\\
  \langle t_A,0\rangle & \mbox{if } \delta_A(p, e)=t_A.
\end{array} \right.
\end{eqnarray*}
As we mentioned in last proof, all the states starting with $ t_A$
are equivalent. Thus, we replace them with one sink state: $\langle
t_A,0\rangle$. Clearly, $B$ accepts the language $L(M)^R\cap L(N)$
and it has $2^m\cdot n-n+1$ states. Next we prove that $B$ is a
minimal DFA.
\begin{itemize}
\item[{\rm (I)}]Every state of $B$ is reachable from $\langle
\{0\},0\rangle$.

Let $\langle p,q\rangle$ be an arbitrary state of $B$. Then there
always exist a string $d^qw$ such that $\delta_B(\langle
\{0\},0\rangle, d^qw) = \langle p,q\rangle$, where $w\in
\{a,b,c\}^*$ and $\delta_A(\{0\},w)=p$.

\item[{\rm {II}}]Any two different states $\langle p_1,q_1\rangle$ and $\langle p_2,q_2\rangle$ of $B$ are distinguishable.
\begin{itemize}
\item[{\rm 1.}]$q_1=q_2$.

In this case, we can find a string $d^iw$ such that
\begin{eqnarray*}
\delta_B(\langle p_1,q_1\rangle, d^iw)& \in & F_B,\\
\delta_B(\langle p_2,q_2\rangle, d^iw) & \notin & F_B,
\end{eqnarray*}
where $i+q_1\mbox{ mod }n= 0$, $w\in \{a,b,c\}^*$, $\delta_A(p_1,
w)\in F_A$ and $\delta_A(p_2, w) \notin F_A$.

\item[{\rm 2.}]$p_1=p_2$, $q_1\neq q_2$.

A string $d^{n-q_1}w$ can distinguish states $\langle
p_1,q_1\rangle$ and $\langle p_2,q_2\rangle$, where $w\in
\{a,b,c\}^*$ and $\delta_A(p_1, w)\in F_A$, because
\begin{eqnarray*}
\delta_B(\langle p_1,q_1\rangle, d^{n-q_1}w)& \in & F_B,\\
\delta_B(\langle p_2,q_2\rangle, d^{n-q_1}w) & \notin & F_B.
\end{eqnarray*}

\item[{\rm 3.}]$p_1\neq p_2$, $q_1\neq q_2$.

Since $A$ is a minimal DFA and $p_1\neq p_2$, there always exists a
string $w\in \{a,b,c\}^*$ such that $\delta_A(p_1, w)\in F_A$ and
$\delta_A(p_2, w) \notin F_A$. Then it is clear that
\begin{eqnarray*}
\delta_B(\langle p_1,q_1\rangle, d^{n-q_1}w)& \in & F_B,\\
\delta_B(\langle p_2,q_2\rangle, d^{n-q_1}w) & \notin & F_B.
\end{eqnarray*}

\end{itemize}
\end{itemize}
Now we know DFA $B$ is minimal because all the states in $B$ are
reachable and distinguishable. Thus, any DFA accepting $L(M)^R\cap
L(N)$ needs at least $2^m\cdot n-n+1$ states. $\ \ \Box $

Theorem~\ref{reversal intersection lower bound} gives a lower bound
of state complexity of $L(M)^R\cap L(N)$. It coincides with the
upper bound shown in Theorem~\ref{reversal intersection upper
bound}. So we have the following theorem.
\begin{theorem}
\label{Tight bound of reversal intersection}

For any integer $m\geq 2$, $n\geq 2$, $2^m\cdot n-n+1$ states are
both sufficient and necessary in the worst case for a DFA to accept
$L(M)^R\cap L(N)$, where $M$ is an $m$-state DFA and $N$ is an
$n$-state DFA.
\end{theorem}

\section{Conclusion}\label{sec:conclusion}

In this paper, we have studied the state complexities of union and
intersection combined with star and reversal. We have proved the
state complexities of four particular combined operations,
including: $L(M)^*\cup L(N)$, $L(M)^*\cap L(N)$, $L(M)^R\cup L(N)$
and $L(M)^R\cap L(N)$ for $m \ge 2$ and $n\ge 2$. They are less than
the mathematical composition of state complexities of its component
operations. The state complexities of the four combined operations
are also less than the state complexities of the combined operations
composed of the same individual operations but in different orders.
The reason of this is that the state complexities decrease when we
perform union and intersection in the end instead of star or
reversal. This makes the order of the state complexities reduced
from $O(2^{m+n})$ to $O(2^{m}n)$. An interesting question is: why
are the state complexities of $L(M)^*\cup L(N)$ and $L(M)^*\cap
L(N)$ the same whereas the state complexities of $(L(M)\cup L(N))^*$
and $(L(M)\cap L(N))^*$ are different?

\end{document}